\documentclass[aps,pra,reprint,superscriptaddress,floatfix]{revtex4-2}
\usepackage[T1]{fontenc}
\usepackage[utf8]{inputenc}
\pdfoutput=1 
\usepackage{graphicx}
\usepackage{amsmath}
\usepackage{amssymb}
\usepackage{bm}
\usepackage{physics}
\renewcommand{\Im}{\operatorname{Im}}
\usepackage[colorlinks, allcolors={Blue}]{hyperref}

\usepackage[capitalize]{cleveref}
\crefrangeformat{equation}{Eqs.~(#3#1#4)--(#5#2#6)} 
\usepackage[caption=false]{subfig}
\newcommand{\phantomsubfloat}[1]{
    {%
        \captionsetup[subfigure]{labelformat=empty}
        \subfloat[][]{#1}
    }%
}
\usepackage[dvipsnames]{xcolor}
\usepackage{orcidlink}

\newcommand{\mvphantom}[1]{\raisebox{0ex}[#1][0ex]{}}

\newcommand{\appsection}[1]{\section{\MakeUppercase{#1}}} 

\begin{document}

\author{Piper Fowler-Wright\orcidlink{0000-0003-1060-445X}}
\affiliation{SUPA, School of Physics and Astronomy, University of St Andrews, St Andrews, KY16 9SS, United Kingdom}
\author{Krist\'{i}n B. Arnard\'{o}ttir\orcidlink{0000-0002-6624-2307}}
\affiliation{SUPA, School of Physics and Astronomy, University of St Andrews, St Andrews, KY16 9SS, United Kingdom}
\author{Peter Kirton\orcidlink{0000-0002-3915-1098}}
\affiliation{Department of Physics and SUPA, University of Strathclyde, Glasgow, G4 0NG, United Kingdom}
\author{Brendon W. Lovett\orcidlink{0000-0001-5142-9585}}
\affiliation{SUPA, School of Physics and Astronomy, University of St Andrews, St Andrews, KY16 9SS, United Kingdom}
\author{Jonathan Keeling\orcidlink{0000-0002-4283-552X}}
\affiliation{SUPA, School of Physics and Astronomy, University of St Andrews, St Andrews, KY16 9SS, United Kingdom}

\title{Determining the validity of cumulant expansions for central spin models}
\date{\today} 
\begin{abstract}
	For a model with many-to-one connectivity it is widely expected that
	mean-field theory captures the exact many-particle \(N\to\infty\) limit, and that higher-order
	cumulant expansions of the Heisenberg equations converge
 to this same limit whilst providing improved approximations at finite \(N\).
 Here we
	show that this is in fact not always the case. Instead, whether mean-field
	theory correctly describes the large-\(N\) limit depends on how the model parameters scale with \(N\), and the convergence of cumulant expansions may be
	non-uniform across even and odd	orders.  Further, even when a higher-order
	cumulant expansion does recover the correct limit, the error is not
	monotonic with \(N\) and may exceed that of mean-field theory.
 
	\vspace{0.8\baselineskip}
	\noindent
	\small{DOI:} \href{https://doi.org/10.1103/PhysRevResearch.5.033148}{10.1103/PhysRevResearch.5.033148}
\end{abstract}

\maketitle

\section{Introduction}

Networks in which one site couples non-locally to many satellite sites
occur in a wide range of many-body open quantum systems.  For example, models
where a driven electronic spin interacts with a bath of nuclear spins
are relevant to 
nuclear magnetic resonance 
spectroscopy~\cite{lilly2017,fernandez2018,villazon2021,rizzato2022},
quantum sensing~\cite{schirhagl2014,wu2016,allert2022} 
and quantum information processing~\cite{taylor2003,divincenzo2005,koppens2006,hanson2007,maurer2012,childress2013,cai2013}. The network structure is also common in
quantum optics where it defines the interaction of a bosonic mode
with an ensemble of emitters~\cite{kirton2019}, or equally a 
single emitter with many
electromagnetic modes~\cite{sanchez2020}.
In many such cases, the large
number of satellite sites 
precludes exact calculations, particularly when accounting for non-unitary dynamics due to incoherent processes. 
Consequently there is a need for approximate methods capable of handling large,
driven-dissipative systems with many-to-one connectivity.
We discuss below how mean-field theory and cumulant expansions may provide a suitable set of methods.

For models with finite connectivity,
mean-field theory is typically only accurate in high dimensions~\cite{chaikin1995}. In
contrast, there are many reasons to believe it should recover the exact behavior of many-to-one models in the thermodynamic limit. First,
given \(N\) identical satellites, monogamy of entanglement~\cite{osborne2006}
restricts the entanglement between any two sites such that
quantum correlations in the system vanish as \(N\to\infty\). However, there is
no similar restriction on \textit{classical} correlations which may certainly
persist in this limit.
Second, in models with weak couplings to satellite sites, these may be
treated as a harmonic bath for the central site with a linear response
that becomes exact as \(N\to\infty\)~\cite{makri1999a}.
Third, for models with an interaction between a large number of emitters and a
bosonic mode, the mean-field equations can be justified via saddle-point
analysis~\cite{eastham2001}. There are further rigorous results regarding the
exactness of mean-field theory as \(N\to\infty\) within this class~\cite{mori2013,carollo2021,fiorelli2023}.
In spite of these results, we present here a simple example where 
mean-field theory does not always capture the \(N\to\infty\) limit of a many-to-one model.

Even when mean-field theories correctly describe the exact $N\to\infty$
		behavior, other methods may be required to capture effects at finite
		\(N\).  Different forms of cumulant expansion of the Heisenberg
		equations have been widely applied to many-body
systems~\cite{kubo1962,fricke1996,vardi2001,kohler2002,kira2008,kramer2015,
kirton2018,reiter2020,sanchez2020,arnardottir2020,robicheaux2021,wang2021,
plankensteiner2022,kusmierek2022,werren2022,huang2022,rubies-bigorda2023} as a
systematic approximation scheme in which increasing orders of correlations are
included; this is hoped to improve accuracy at the cost of growing complexity.
The power of this approach is the small dimension of the resulting problem
(independent of system size \(N\)), and the ability of even low orders of
expansion to produce accurate results at intermediate \(N\). Hence, they
are a tool to both capture behavior at \(N\gg 1\) and to study finite size
effects.

The difficulty of direct simulation at large \(N\) 
means that cumulant expansions are rarely
benchmarked against exact methods much beyond \(N\sim 30\).  Confidence in
results may then be based on the assumptions that evaluation at larger \(N\) and
higher orders of expansion provide more accurate approximations.  However, we
show cases here where neither of these assumptions are correct. 

In this work we thoroughly explore the convergence of cumulant expansions for
a driven-dissipative central spin model.  We demonstrate how the ability of
		mean-field theory to capture the \(N\to\infty\) steady state of the full
		quantum model depends on the scaling of parameters. Further, we show how
		even when mean-field theory does capture the exact behavior at
		\(N\to\infty\), convergence of higher-order cumulant expansions to the
		same result is not guaranteed. We discuss how this convergence
		behavior arises in light of correlations present in the system and show
		that similar behavior may be observed in models of light-matter
		interaction.  Permutation symmetry allows us to make comparisons to
		exact results for the central spin model at relatively large \(N\sim
		150\) whereby we show the error in cumulant expansion approximations
		does not generally decrease monotonically with \(N\), nor with the order of
		expansion. 

The structure of the paper is as follows. In \cref{sec:model} we give an overview
of the central spin model and the permutation symmetric method that may
be used to solve it at finite \(N\). In \cref{sec:mean-field_cumulant} we
explain the cumulant expansion method and its application to the model at
mean-field and second order. \Cref{sec:results} then compares the results for
these approximations up to third order to exact data under two different 
choices for scaling of parameters as \(N\to\infty\). Finally, in 
\cref{sec:higher-order} we  present the results for higher-order expansions in 
both the central spin and the Tavis--Cummings models before summarizing our 
findings and the scope for future work in \cref{sec:conclusion}.

\section{Model}
\label{sec:model}

\begin{figure}
	\centering
    \vspace{-2\baselineskip}%
	\phantomsubfloat{\label{fig:1a}}
    \phantomsubfloat{\label{fig:1b}}

	\includegraphics[width=\linewidth]{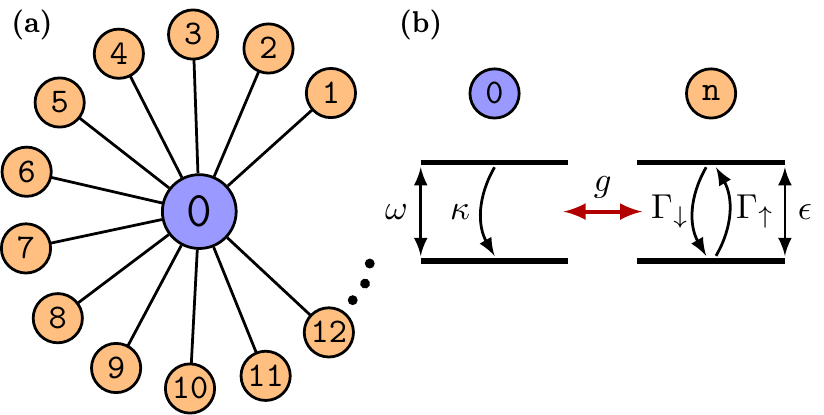}%
	\caption{%
		(a) Network of the model: a central site (index \(0\)) couples to
		\(N\) identical satellites (\(n=1,\ldots, N\)). 
		(b) Each site is a two-level system (spin-1/2)
		subject to decay (\(\kappa\)
		or \(\Gamma_\downarrow\)) and, in the
		case of the satellites, pump \(\Gamma_\uparrow\).
}
\label{fig:1A}
\end{figure}

We consider a single spin-1/2 (Pauli matrices \(\sigma^\alpha_0\))
interacting with \(N\)  spin-1/2 satellites (Pauli matrices \(\sigma^\alpha_n\)) according to
\begin{equation}
	H= \frac{\omega}{2}\sigma^z_0 +  \sum_{n=1}^N\left[\vphantom{\frac{\Omega}{\Omega}}
		\frac{\epsilon}{2}\sigma^z_n + g\left( \sigma^+_0 \sigma^-_n
	+ \sigma^-_{0} \sigma^+_{n}\right)\right].
	\label{eq:H}
\end{equation}
Here \(\omega\) and \(\epsilon\) are  on-site energies for the central and
a satellite spin, and \(g\) the interaction strength.
In addition we consider dissipation with rate \(\kappa\) from the central site
as well as incoherent pump \(\Gamma_\uparrow\) and loss \(\Gamma_\downarrow\)
for each satellite. These are included as Markovian terms in the master equation
for the total density operator \(\rho\),
\begin{equation}
\partial_t \rho	 = - i \left[ H, \rho \right]
+  \kappa \mathcal{L}[\sigma^-_{0}]
	+
	 \sum_{n=1}^N \bigl(\Gamma_\uparrow \mathcal{L}[\sigma^+_n]
	 +  \Gamma_\downarrow \mathcal{L}[\sigma^-_n]\bigr)\text{,}
	 \label{eq:ME}
\end{equation}
with \(\mathcal{L}[x]=x\rho x^\dagger - \{x^\dagger x, \rho\}/2\). Schematics
for the system and these processes are given in  \cref{fig:1a,fig:1b}.

The anisotropic interactions in \cref{eq:H}
arise, for example, between the nitrogen-vacancy
center and the \({}^{13}\text{C}\) nuclear spins in diamond~\cite{doherty2013}.
This system has been
extensively studied for its potential role
in emerging quantum technologies including 
 spectroscopy~\cite{fernandez2018,villazon2021,rizzato2022},
 quantum sensing~\cite{schirhagl2014,wu2016,allert2022}, and
computing~\cite{maurer2012,childress2013}.
For our purpose the model serves a minimal formulation of the 
open many-to-one problem to investigate mean-field theory and cumulant expansions.
In certain cases, such as the absence of
dissipation, or when the satellite dissipation
is collective, there exist analytical or efficient numerical methods
capable of accessing large-\(N\) behavior of central spin models~\cite{chen2007,lindoy2018,ribeiro2019,ricottone2020,villazon2020,malz2022}. However,
for the case we consider with individual dephasing these methods do not apply.

The model~\cref{eq:ME}
has cumulant equations that are analytically
tractable up to third order whilst also allowing exact calculations
for relatively large system sizes.
Below, to compare approximations, we analyze the central-site population, \(p^\uparrow_0\), in the steady state. This relates to the polarization, 
\(\langle \sigma^z_{0} \rangle\), via
\(p^\uparrow_0\equiv(1+\langle \sigma^z_{0} \rangle)/2\) and increases from zero as the ratio \(\Gamma_\uparrow/\Gamma_T\) (\(\Gamma_T=\Gamma_\uparrow+\Gamma_\downarrow\))
is increased.

The invariance of the model under the interchange of satellite spins
allows one to work in a permutation symmetric basis when performing exact
calculations~\cite{chase2008,xu2013,damanet2016,gegg2016,kirton2017,shammah2018}. This provides a combinatoric reduction in the
size of the Liouvillian \(L\). In our case this allows finding
the eigenvector of \(L\) with eigenvalue \(0\), i.e., the steady state, up to \(N=150\). 
No information is lost by working in this basis. In particular, all
correlations can be computed exactly and compared to the prediction of the
cumulant expansions.

\section{Mean-field and cumulant expansions}
\label{sec:mean-field_cumulant}
\begin{figure}
	\centering
    \vspace{-2\baselineskip}%
	\phantomsubfloat{\label{fig:1c}}
    \phantomsubfloat{\label{fig:1d}}

	\includegraphics[width=\linewidth]{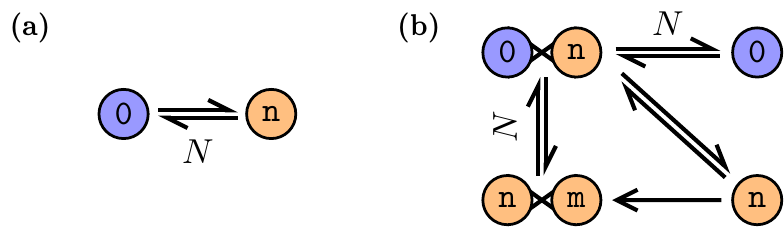}%
	\caption{%
		(a)
		Mean-field reduction to a two-body problem where expectations of a
		satellite  evolve according to expectations of the central site 
  		(\scalebox{1.5}{\raisebox{-.25ex}{\(\bm{\rightharpoonup}\)}})
		which in turn evolve according to \(N\) copies of the satellite expectations
		(\scalebox{1.5}{\raisebox{-.25ex}{\(\bm{\leftharpoondown}\)}}).
		(b) In the second-order cumulant expansion central-satellite
		and satellite-satellite expectations couple into the system
		[\cref{eq:c2a,eq:c2b}].
}
\label{fig:1B}
\end{figure}

We now explain the cumulant expansion  method and its application to 
the central spin model at mean-field
and second order. Expressions for third-order cumulant equations 
are also provided in \cref{app:1}.  

From the master equation, \cref{eq:ME}, one can derive
equations of motion for single-site expectations,
\begin{align}
	\partial_t \langle \sigma^z_{0} \rangle &= - \kappa\left( \langle \sigma^z_{0} \rangle+1 \right)
	+ 4 g N \Im\!\left[\langle \sigma^+_{0}\sigma^-_{n} \rangle\right],
	\label{eq:mfa}\\
	\partial_t \langle \sigma^z_{n} \rangle &=  - \Gamma_T \langle \sigma^z_{n} \rangle + \Gamma_\Delta
	- 4 g  \Im\!\left[\langle \sigma^+_{0}\sigma^-_{n} \rangle\right],
	\label{eq:mfb}\\
	\partial_t \langle \sigma^+_{0} \rangle &=
	\left(\mvphantom{2.6ex}i\omega-\smash{\frac{\kappa}{2}}\right) \langle \sigma^+_{0} \rangle - igN
	\langle\sigma^z_{0} \sigma^+_{n} \rangle, \label{eq:mfc}\\
	\partial_t \langle \sigma^+_{n} \rangle &=
	\left(\mvphantom{2.9ex}i\epsilon-\smash{\frac{\Gamma_T}{2}}\right) \langle \sigma^+_{n} \rangle - ig
	\langle\sigma^+_{0} \sigma^z_{n} \rangle, \label{eq:mfd}
\end{align}
where \(\Gamma_\Delta=\Gamma_\uparrow-\Gamma_\downarrow\) and
\(\Gamma_T=\Gamma_\uparrow+\Gamma_\downarrow\).  This set of equations is not
closed since, for example, \(\partial_t\langle \sigma^z_{0} \rangle\) depends on
\(\langle \sigma^+_{0} \sigma^-_{n} \rangle\).  The equation for \(\langle
\sigma^+_{0} \sigma^-_{n} \rangle\) will in turn depend on expectations of
operators from three different sites, and so on, resulting in an exponential
number of equations involving operators on all sites.

To obtain a manageable number of equations,  in the \(M^\text{th}\)-order
cumulant
expansion moments of order \(M+1\) are rewritten as  non-linear combinations of
lower-order moments by setting the corresponding
\textit{cumulant}~\cite{gardiner2009,kubo1962} to zero.  Such an approximation
corresponds to making an ansatz for the many-body state \(\rho\) that involves correlations between at most \(M\) sites. We stress here the distinction is between sites
(or Hilbert spaces) of the many-body system, not operators in themselves (as,
e.g., used in Ref.~\cite{kira2008}).  This is natural for two-level systems,
where one easily identifies \(\langle \sigma^+_{0}\sigma^{-}_0\sigma^z_{n}
\rangle= (\langle \sigma^z_{0}\sigma^z_{n} \rangle
+\langle\sigma^z_n\rangle)/2\), but for bosonic operators, e.g., \(a\), it is
common to see factorizations such as \(\langle a^\dagger a \sigma^z_{} \rangle
\approx \langle a^\dagger \rangle \langle a \rangle \langle \sigma^z_{}
\rangle+\ldots\) whose validity depends on additional assumptions of
Gaussianity~\cite{weedbrook2012}.

\subsection{Mean-field equations}
At first order, that is mean-field theory, second-order moments factorize into
products
(\(\langle \sigma^\alpha_{0} \sigma_n^\beta \rangle\approx \langle \sigma^\alpha_{0}\rangle\langle \sigma^\beta_{n} \rangle\)) and an effective two-body problem results [\cref{fig:1c}]. Solving for the steady state one finds 
\(\langle \sigma^z_{0} \rangle=-1\) for 
\(\Gamma_\uparrow/\Gamma_T\) below a critical pump ratio 
\(R_c\equiv(1+\Gamma_T\kappa/4g^2N)/2\), while for \(\Gamma_\uparrow/\Gamma_T>R_c\):
\begin{align}
\begin{split}\hspace{-.7cm}
	\langle \sigma^z_{0} \rangle = &- \frac{1}{2}\left( 1- \frac{\Gamma_\Delta N}{\kappa} \right)- \frac{1}{2}
\sqrt{
\left( 1- \frac{\Gamma_\Delta N}{\kappa} \right)^2
+
\frac{\Gamma_T^2}{g^2} 
}\text{,}\hspace{-.25cm}
\end{split}\label{eq:mf_sola}\\
\langle \sigma^z_{n} \rangle &= -\frac{\kappa\Gamma_T}{4g^2N\langle \sigma^z_{0} \rangle},\quad
\langle \sigma^+_{n} \rangle = \frac{i\kappa}{2gN \langle \sigma^z_0\rangle} \langle \sigma^+_{0} \rangle,\label{eq:mf_solb}
\end{align}
where the magnitude of \(\langle \sigma^+_{0} \rangle\) is fixed by
\begin{align}
	\abs{\langle \sigma^+_{0} \rangle}^2 = - \langle \sigma^z_{0} \rangle\left(1+\langle \sigma^z_{0} \rangle\mvphantom{2ex}\right)\!/2.\label{eq:mf_solc}
\end{align}
For simplicity we took \(\omega=\epsilon\) above but have checked our conclusions do not change off resonance.

Although the model has \(\text{U}(1)\) symmetry, i.e.,  \cref{eq:ME}  is invariant
under \(\sigma^{\pm}\to \sigma^{\pm} e^{\pm i \theta}\), it is necessary to
retain the symmetry-breaking terms \(\langle \sigma^+_0 \rangle\) and \(\langle
\sigma^+_n \rangle\) when performing the mean-field approximation in order to
obtain a non-trivial solution:  the state \(\langle \sigma^z_0 \rangle = -1\) is
always a solution to the mean-field equations that only becomes unstable when \(\Gamma_\uparrow/\Gamma_T >
R_c\).

\subsection{Second-order cumulant equations}
Breaking symmetry is not necessary at second order where
\(\langle \sigma^+_0 \sigma^-_n\rangle\) can be non-zero whilst respecting the symmetry. 
The required equations for second moments are [\cref{fig:1d}]
\begin{align}
\partial_t \langle  \sigma^+_{0} \sigma^-_{n} \rangle &= 
\left(\mvphantom{2.9ex}i(\omega-\epsilon) -\smash{\frac{\kappa+\Gamma_T}{2}} \right) \langle \sigma^+_{0}\sigma^-_{n} \rangle
+ \frac{ig}{2} \langle \sigma^z_{n} \rangle \nonumber\\
&-	\frac{ig}{2} \langle \sigma^z_{0} \rangle 
	- ig(N-1) \langle \sigma^z_{0}\rangle\langle  \sigma^+_{n}\sigma^-_{m} \rangle, 
	\label{eq:c2a}
\\
	\partial_t\langle \sigma^+_{n}\sigma^-_{m} \rangle &=
		- \Gamma_T\langle \sigma^+_{n}\sigma^-_{m} \rangle
  +2g\Im\!\left[\langle \sigma^+_{0} \sigma^-_{n}\rangle\right] \langle\sigma^z_{n}\rangle,
\label{eq:c2b}
\end{align}
with \(n\neq m\).
Here we set third cumulants to zero and use the  \(\text{U}(1)\) symmetry  to write
\(\langle\sigma^z_{0} \sigma^+_{n}\sigma^-_{m}  \rangle\approx\langle
\sigma^z_{0}\rangle\langle  \sigma^+_{n}\sigma^-_{m} \rangle\),
\(\langle\sigma^+_{0} \sigma^-_{n}\sigma^z_{m}  \rangle\approx\langle
\sigma^+_{0}\sigma^-_{n} \rangle \langle \sigma^z_{n} \rangle\).
\Cref{eq:mfa,eq:mfb,eq:c2a,eq:c2b} can also be solved exactly, albeit not
explicitly, to find \(p^\uparrow_0=[1+\langle \sigma^z_0
\rangle]/2\). 

\section{Results at mean-field and second-order cumulants} 
\label{sec:results}
In the following we compare the mean-field
result \cref{eq:mf_sola} and the solution to the second-order
equations \cref{eq:mfa,eq:mfb,eq:c2a,eq:c2b} to the exact steady state.  We do this this under two
possible choices for scaling parameters in the model as \(N\to \infty\).

\subsection{Fixed \texorpdfstring{\(g\sqrt{N}\)}{gSqrtN}}
\begin{figure}
	\centering
    \vspace{-2\baselineskip}%
	\phantomsubfloat{\label{fig:2a}}%
    \phantomsubfloat{\label{fig:2b}}%
    \phantomsubfloat{\label{fig:2c}}%
    \phantomsubfloat{\label{fig:2d}}%

	\hspace{-.4cm}%
	\includegraphics[width=\linewidth]{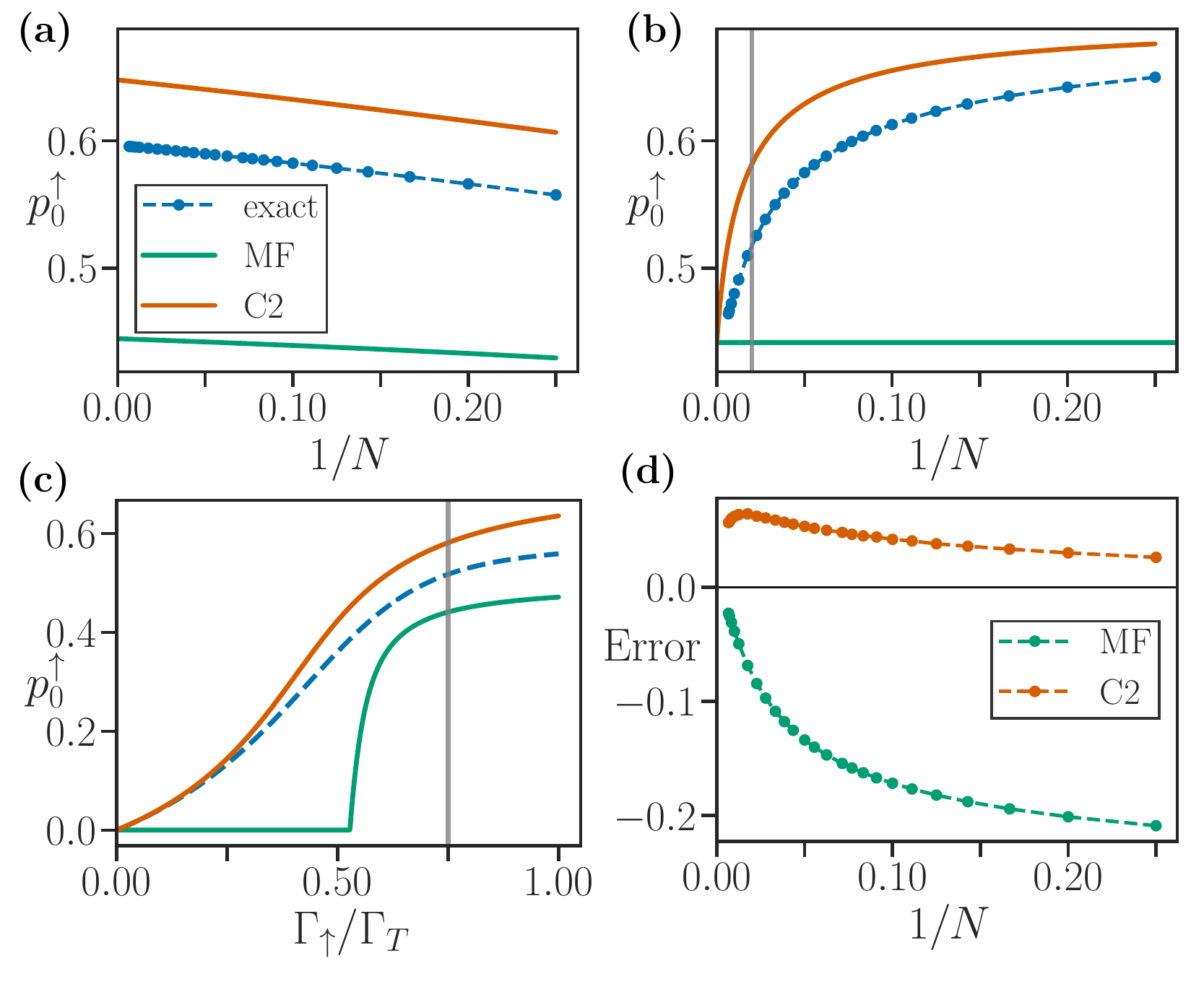}%
	\caption{%
		(a) Central-site population, \(p^\uparrow_0 =[1+\langle \sigma^z_{0} \rangle]/2\),
		in mean-field (MF) and second-order cumulant (C2) approximations when
		\(g\sqrt{N}=3\) is fixed and \(\kappa=1\) in units of \(\omega\). Exact data (blue dots) is included
		up to \(N=150\).
        The horizontal scale \(1/N\) is such that \(N\to\infty\)
        to the left. %
		(b) MF, C2 and exact results when \(\kappa/N=1/16\) is fixed and
		\(g=3/4\). The gray vertical line at \(N=50\) indicates points equivalent to those along the corresponding
		line in (c).
		(c) \(p^\uparrow_0\) vs  \(\Gamma_\uparrow/\Gamma_T\) (\(\Gamma_T=\Gamma_\uparrow+\Gamma_\downarrow\))
		at fixed \(\kappa/N=1/16\), \(g=3/4\), and \(N=50\)
		(the low cost of the exact calculation allowed a continuous line to be plotted).
        The mean-field transition
        at  \(\Gamma_\uparrow/\Gamma_T=R_c\approx0.53\)
        is analogous to that
        in a driven-dissipative Tavis--Cummings model~\cite{kirton2019} and
        other models of lasing~\cite{werren2022,frantzeskakis2023}.
		(d) Error in MF and C2 results from (b). 
		Other parameters used in these panels were \(\epsilon=\omega=1\), \(\Gamma_T=2\),	and [except (c)] \(\Gamma_\uparrow=3/2\).}
	\label{fig:2}
\end{figure}

Figure~\ref{fig:2a} shows \(p^\uparrow_0\) vs \(1/N\)
when fixing \(g\sqrt{N}\).
This scaling is often relevant in the context of light-matter coupling, where coupling strength \(g\) is inversely proportional to the square root of mode volume: as the system becomes larger,  both \(N\) and mode volume grow,  but \(g\sqrt{N}\) remains fixed.
Here we see there is no agreement between the exact and approximate results,
each taking different \(N\to\infty\) limits.  This is in marked contrast to the
Tavis--Cummings or Dicke models~\cite{kirton2017}, 
where both mean-field and second-order cumulant approximations 
converge to the exact steady-state as \(N\to\infty\)  for this scaling.
Below we explain how the convergence of second-order cumulants to mean-field theory is precluded by
\(g\propto 1/\sqrt{N}\) for the central spin model.

\subsection{Fixed \texorpdfstring{\(\kappa/N\)}{kappa/N}}
\label{sub:fixed_kappa}
If instead the ratio \(\kappa/N\) is kept fixed, \cref{fig:2b}, mean-field and second-order cumulants have
a common limit that captures the exact behavior.
Note \cref{fig:2b} is plotted for parameters where non-zero \(p^\uparrow_0\) is
expected; see \cref{fig:2c} for a phase diagram.
This scaling may be understood to realize the limit of strong continuous 
measurement of the central site~\cite{krishna2023}.
It has the feature,
seen in \crefrange{eq:mf_sola}{eq:mf_solc}, that expectations of satellite and
central-site quantities are of the same order, \(O(1)\), as \(N\to\infty\). In
\cref{app:2} we show this holds for higher-order correlations as well.  One then
observes the asymptotic form of \cref{eq:c2a} (\(\kappa\sim N\)),
\begin{align} 
\begin{split}
	\partial_t \langle \sigma^+_{0} \sigma^-_{n} \rangle	
	&=  N \left(-\frac{\kappa}{2N} \langle \sigma^+_{0}\sigma^-_{n} \rangle
- ig \langle \sigma^z_{0}\rangle\langle  \sigma^+_{n}\sigma^-_{m} \rangle\right) \\
&+ O(1),
\end{split}
\end{align}
matches that predicted by mean-field theory,
\begin{align}
	 \begin{split}
 \partial_t \left(\langle \sigma^+_{0}\rangle\langle \sigma^-_{n} \rangle  \right) 
 &=  N\left(-\frac{\kappa}{2N} \langle \sigma^+_{0}\rangle\langle\sigma^-_{n} \rangle
 \right.\\&\hspace{.9cm}\left.\vphantom{\frac{\kappa}{2}}
 - ig \langle \sigma^z_{0}\rangle\langle  \sigma^+_{n}\rangle\langle\sigma^-_{m} \rangle\right)+O(1).
  \end{split}
\end{align}
The same is true for \cref{eq:c2b} and its mean-field analog, hence the second-order and
mean-field equations have identical structures as \(N\to\infty\) at fixed \(\kappa/N\). 

In contrast at fixed \(g\sqrt{N}\) the correlations \(\langle \sigma^+_n
\sigma^-_m \rangle \) do not remain finite as \(N\to\infty\) but decay faster
than \(1/\sqrt{N}\) (\cref{app:2}).  Consequently the terms \(\sim g \langle
\sigma^z_0\rangle,g\langle \sigma^z_n \rangle\) in \cref{eq:c2a}, which are
not present in mean-field theory, cannot be discounted as \(N\to\infty\).  This
difference leads to distinct limits in \cref{fig:2a}. Note equations for
higher-order moments involving the central site will contain additional terms
inconsistent with mean-field theory. Thus, while higher-order expansions
\textit{may} provide an improved approximation of the exact results, they will
generally have distinct limits.  This result also illustrates how knowledge that
certain correlations vanish at large \(N\) is  not  sufficient to determine if
they become irrelevant as \(N\to\infty\).  Instead, the scaling with \(N\) of
parameters multiplying these correlations must also be taken into account.

 \cref{fig:2d} shows further how with \(\kappa/N\) fixed the error at second order is not monotonic with \(N\) 
and even exceeds that of mean-field theory for \(N\gtrsim80\). 
The non-monotonicty is inevitable given this approach 
captures the exact \(N\to\infty\) limit and must also be exact at \(N=2\), 
when all correlations are fully captured. 
As such, the second-order  expansion provides an approximation that is only asymptotically matched
to the exact result at the two limits, and care must be taken in between.

\section{Higher-order cumulant expansions}
\label{sec:higher-order}
\subsection{Central spin model}

\begin{figure}[t]
	\centering
    \vspace{-2\baselineskip}%
	\phantomsubfloat{\label{fig:3a}}%
    \phantomsubfloat{\label{fig:3b}}%

	\hspace{-.4cm}%
	\includegraphics[width=\linewidth]{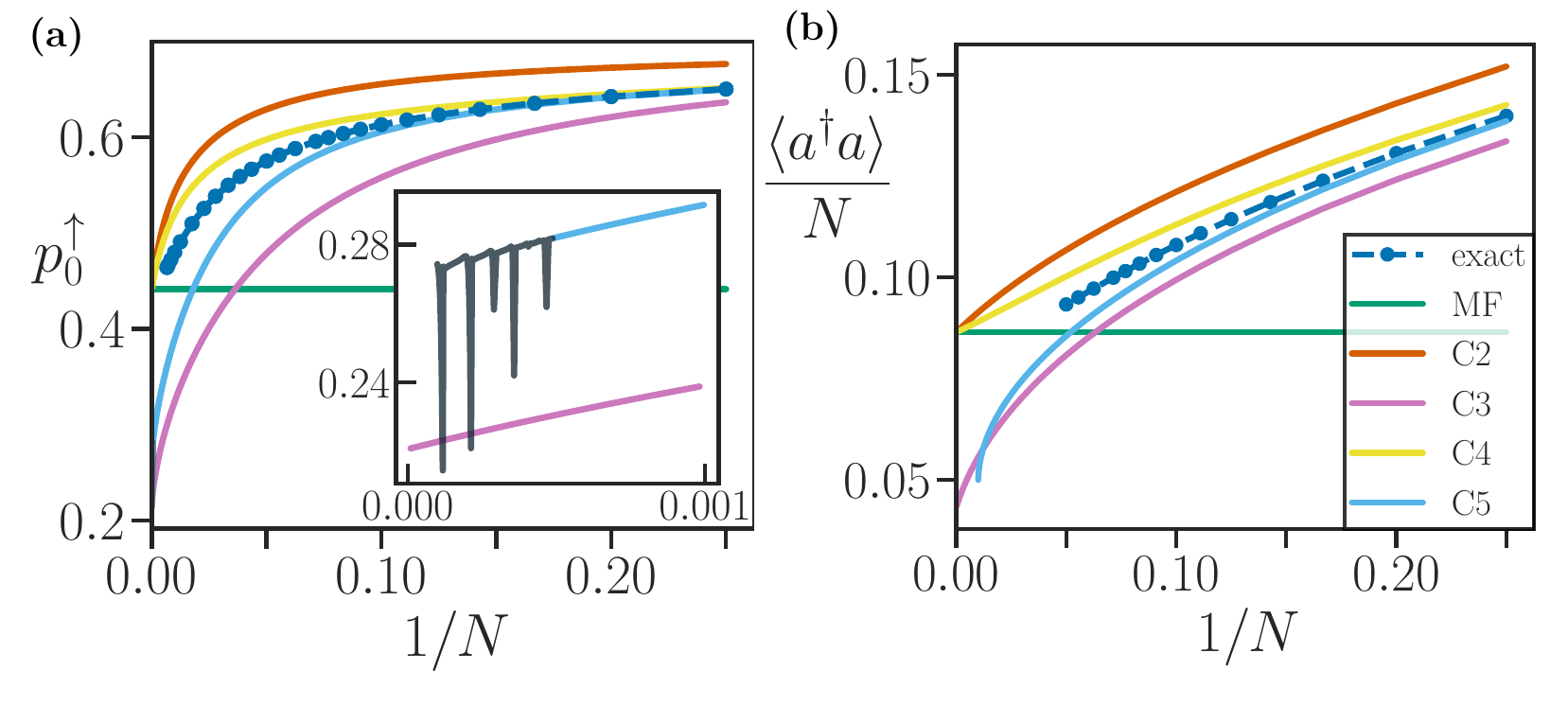}%
	\caption{%
		(a) Central-site population \(p^\uparrow_0 \)
		in mean-field (MF) and cumulant approximations up to fifth order (C2-5) at fixed \(\kappa/N=1/16\)
		[parameters and exact data as in \cref{fig:2b}]. Results at fourth
		and fifth order were derived using 
        QuantumCumulants.jl~\cite{plankensteiner2022}.
		Inset: the fifth-order solution has numerical noise beyond 
            \(N\gtrsim2,500\), but 
		is approaching a value distinct from the third-order limit.
		(b) Mean-field and cumulant	results for the scaled photon number
		\(\langle a^\dagger a \rangle/N\) in the driven-dissipative Tavis--Cummings model 
        using QuantumCumulants.jl. Exact results following a
        Fock-space truncation are included up to \(N=20\) (\(N_\text{phot.}=20\) levels were sufficient to achieve convergence).
  Here the parameters
        were \(g\sqrt{N}=9/10\),
		\(\epsilon=\omega=\kappa=1\), \(\Gamma_T=1/2\), and \(\Gamma_\uparrow=3\Gamma_T/4\).
  The fifth-order solution became unstable for \(N\gtrsim100\).	
	}
	\label{fig:3}
\end{figure}

Having established a well defined limit up to second order at fixed \(\kappa/N\), 
we now investigate higher-order cumulant expansions for this scaling. We use the  
QuantumCumulants.jl Julia framework~\cite{plankensteiner2022} to obtain
fourth and fifth-order results in addition to the solution to the third-order equations presented in \cref{app:1}. Surprisingly, we see in \cref{fig:3a}
that whilst the fourth-order expansion provides an improved
approximation on the entire range of \(N\), the third-order expansion does not.
Instead it converges to a limit far separated from the
true result, hence there is some \(N\) beyond which the second-order (and mean-field)
result provides a better approximation. Similarly the fifth-order result,
despite being exact 
up to \(N=5\) and the best approximation at very small \(N\),
fails to capture the exact \(N\to\infty\) limit.

To understand the dependence of convergence on order parity, the previous
argument for the asymptotic reduction of the second-order equations to
mean field as \(N\to\infty\) can be extended to all even orders.  First, note
that before any factorization is made the equations for moments involving
satellite sites only match mean-field theory in structure since \(H\)
[\cref{eq:H}] is linear in these sites. When the central site is involved, this
is no longer the case. However, the terms that survive as \(N\to\infty\) at
fixed \(\kappa/N\) are those that arise from the commutator of a central
operator with \(\sigma^{+}_0\sigma^-_n\) or \(\sigma^{-}_0\sigma^+_n\) followed
by a sum \(\sim N\) over the satellites.  These terms have the same structure
for both the cumulant equations and mean-field theory.  Second, there is a key
point about the coefficients associated with the cumulant expansion of a given
term. As discussed
further in \cref{app:3}, by definition, the sum of coefficients of the
cumulant expansion of any given term should sum to 1.  However, when some
terms are eliminated because they do not respect the symmetries of the model,
this statement may or may not remain true.  When moments are factorized at even
orders of expansion, the  number of non-vanishing terms under \(\text{U}(1)\) symmetry
does  sum to 1~\footnote{We have checked this explicitly up to
\(14\)th order.}.  As this matches the mean-field prediction for the
number of terms, the asymptotic structure of even-order equations are compatible
with mean-field theory.

On the other hand, closing the equations at odd orders requires factorizing
moments \(\langle \sigma^+_{0}\sigma^-_{n}\sigma^+_{m}\sigma^-_{k}\ldots
\rangle\) involving raising and lowering operators only. These produce a set of
terms with coefficients that do \textit{not} sum to 1.  For example, when
constructing the third-order equations setting the cumulant \(\langle\langle
\sigma^+_{0} \sigma^-_{n} \sigma^+_{m} \sigma^-_{k}\rangle\rangle\) to zero
gives
\begin{align}
	\langle  \sigma^+_{0} \sigma^-_{n} \sigma^+_{m} \sigma^-_{k}  \rangle
	\approx 2 \langle \sigma^+_{0}\sigma^-_{n} \rangle\langle \sigma^+_{m}\sigma^-_{k} \rangle,
	\label{eq:3factor}
\end{align}
having excluded terms that vanish on account of the \(\text{U}(1)\) symmetry. 
It is the factor of \(2\) occurring 
here that is incongruent with mean-field theory.  
The number of terms produced by these type of factorizations varies with
successive odd orders (\(2,-3,34,-455\)\ldots), so
each can be expected to converge on its own limiting value at \(N\to\infty\),
as observed in \cref{fig:3a} for third and fifth orders. 

A consequence of these observations is that symmetry-broken versions
of the odd-order equations, for which no terms of the approximation for 
\(\langle \sigma^+_{0} \sigma^-_{n}\sigma^+_{m}\sigma^-_{k}\ldots \rangle\) vanish,
can produce the correct limit. In \cref{app:3} we 
show this is indeed the case for our model.  However, we note that at finite \(N\) the exact solution never shows symmetry breaking, and that 
the symmetry-broken approximation is not necessarily a reliable improvement.

\subsection{Tavis--Cummings model}
Finally we observe similar  convergence behavior 
between even and odd orders in models of light-matter interaction. Figure~\ref{fig:3b}
includes results for the driven-dissipative 
Tavis--Cummings model up to fifth order of the cumulant expansion~\footnote{
	As mentioned in the paper, working with the cumulant equations by counting operators \(a\)
	requires treating multiple photon operators as distinct objects when breaking
	higher-order moments to obtain a closed set of equations.
	Alternatively, one can consider a truncation of the photon space to \(N_{\text{phot.}}\)
	levels and work, e.g., in the basis
	of generalized Gell-Mann matrices. This increases the number of equations at any order
	of expansion but removes the reliance on additional assumptions, i.e., Gaussianity.
}.
The Tavis--Cummings Hamiltonian~\cite{tavis1968}
is frequently used in cavity QED to describe an ensemble of non-interacting
emitters coupled to a common cavity mode and may be obtained from
\cref{eq:H} by replacing the central spin with a bosonic operator \(a\)~\cite{kirton2019}. Note for this model \(g\sqrt{N}\) fixed
provides matching exact and mean-field
\(N\to\infty\) limits for the steady state~\cite{kirton2017}.

\section{Discussion}
\label{sec:conclusion}
In this paper
we examined the convergence of mean-field and 
cumulant expansions at \(N\to\infty\) as well as
their accuracy at intermediate \(N\).
We considered the class of all-to-one models for which mean-field theory may be expected to be robust. 
Yet for our central spin model 
we demonstrated that whether mean-field theory
captures the exact steady state as \(N\to\infty\) depends on the scaling of parameters in the model. 
Further, even when mean-field theory does capture
exact  \(N\to\infty\) behavior, higher-order cumulant expansions may not converge to the same result.
 Comparison to exact results up to \(N=150\) allowed us to verify
the large-\(N\) behavior and show the error of cumulant expansions
is not monotonic with \(N\). 

The model considered here has been directly applied to
study defect centers in diamond~\cite{fernandez2018,rao2020,villazon2021} and 
quantum 
dot systems~\cite{lai2006,rao2020},
but our reasoning may be applied quite generally to central spin models including,
for example,  other anisotropic or isotropic 
couplings~\cite{al-hassanieh2006,fauseweh2017,rohrig2018,zhou2019}
or coherent drive~\cite{kessler2012,hildmann2014,frantzeskakis2023}.
We have also seen that our results are relevant
to models of collective light-matter coupling where cumulant expansions are an
increasingly popular choice for analyzing both small and large
systems~\cite{kirton2018,sanchez2020,reiter2020,arnardottir2020,robicheaux2021,plankensteiner2022,wang2021,werren2022,kusmierek2022}.

While we focused on steady state properties, 
future work may use the cumulant expansions to examine the dynamics of
open central spin models~\cite{bortz2007,jivulescu2009,stanek2014,zhou2019,bhattacharya2021,tang2022}
for which the scope of mean-field theory to capture exact \(N\to\infty\) behavior
has recently been studied~\cite{fiorelli2023,carollo2023nongaussian}.
Similarly, one may look to apply our reasoning to 
models with all-to-all connectivity considering  
studies~\cite{kramer2015,mahmoodian2020,robicheaux2021,kusmierek2022,rubies-bigorda2023} of the limitations of mean-field approximations in this class.
Our results highlight the need to assess the validity of cumulant expansions
in such applications, and prompt further exploration of how reliable higher-order expansions can be found.

\textit{Note added.} Recently, another work studying this same problem has appeared~\cite{carollo2023nongaussian}.

The research data supporting this publication can be accessed at Ref.~\cite{dataset}.

\begin{acknowledgements}
The authors thank Federico Carollo for valuable discussions.
P.F.-W. acknowledges support from EPSRC (Grant No. EP/T518062/1).
K.B.A, B.W.L., and J.K. acknowledge support from EPSRC 
(Grant No. EP/T014032/1).
\end{acknowledgements}

\appendix

\appsection{Third-order cumulant equations}%
\label{app:1}
In this appendix we provide the third-order cumulant equations for the central spin model with \(\text{U}(1)\) symmetry. 
In the following, \(n\), \(m\), and \(k\) label distinct satellite sites.
\vfill
\begin{widetext}
\vspace{-1.5\baselineskip}
\begin{align}
	\partial_t \langle \sigma^z_{0} \rangle &=
	-\kappa\left( \langle \sigma^z_{0} \rangle+1 \right) + 4gN \Im\!\left[\langle \sigma^+_{0}\sigma^-_{n} \rangle \right]
	\label{eq:c3a}\\
	\partial_t \langle \sigma^z_{n} \rangle &=
	- \Gamma_T\langle \sigma^z_{n} \rangle + \Gamma_\Delta - 4 g\Im\!\left[\langle \sigma^+_{0}\sigma^-_{n} \rangle\right]
	\label{eq:c3b}
	\\
 	\begin{split}
\partial_t \langle  \sigma^+_{0} \sigma^-_{n} \rangle &= 
\left(\mvphantom{2.9ex}i(\omega-\epsilon) -\smash{\frac{\kappa+\Gamma_T}{2}} \right) \langle \sigma^+_{0}\sigma^-_{n} \rangle
+ \frac{ig}{2} \langle \sigma^z_{n} \rangle -	\frac{ig}{2} \langle \sigma^z_{0} \rangle 
	- ig(N-1) \langle \sigma^z_{0} \sigma^+_{n}\sigma^-_{m} \rangle 
	\label{eq:c3c}
\end{split}\\
	\partial_t\langle \sigma^+_{n}\sigma^-_{m} \rangle &=
		- \Gamma_T\langle \sigma^+_{n}\sigma^-_{m} \rangle
  +2g\Im\!\left[\langle \sigma^+_{0} \sigma^-_{n} \sigma^z_{m}\rangle\right]
	\label{eq:c3d}
	\\
	\begin{split}
		\partial_t \langle \sigma^z_{0}\sigma^+_{n}\sigma^-_{m} \rangle &=
  - (\kappa+\Gamma_T) \langle \sigma^z_{0}\sigma^+_{n}\sigma^-_{m} \rangle
	- \kappa\langle \sigma^+_{n}\sigma^-_{m} \rangle 
		+2g \Im\!\left[\langle \sigma^+_{0}\sigma^-_{n} \rangle\right] + 8g(N-2) \Im\!\left[\langle \sigma^+_{0}\sigma^-_{n} \rangle\right]
		\langle \sigma^+_{n}\sigma^-_{m} \rangle 
	\end{split}
	\label{eq:c3e}\\
	\begin{split}
		\partial_t \langle \sigma^+_{0}\sigma^-_{n}\sigma^z_{m} \rangle &=
		\left(\mvphantom{2.9ex}i(\omega-\epsilon) -\smash{\frac{\kappa+3\Gamma_T}{2}} \right) \langle \sigma^+_{0}\sigma^-_{n}\sigma^z_{m} \rangle 
		+ \Gamma_\Delta \langle \sigma^+_{0}\sigma^-_{n} \rangle 
		-ig\langle \sigma^+_{n}\sigma^-_{m} \rangle + \frac{ig}{2} \langle \sigma^z_{n}\sigma^z_{m} \rangle- \frac{ig}{2}\langle \sigma^z_{0}\sigma^z_{n} \rangle \\
		&- ig(N-2) \biggl( \langle \sigma^z_{0}\sigma^+_{n}\sigma^-_{m} \rangle \langle \sigma^z_{n} \rangle
			+ \langle \sigma^z_{0} \rangle\langle \sigma^z_{n}\sigma^+_{m}\sigma^-_{k} \rangle + \langle \sigma^z_{0}\sigma^z_{n} \rangle\langle \sigma^+_{n} \sigma^-_{m} \rangle
- 2 \langle \sigma^z_{0} \rangle\langle \sigma^z_{n} \rangle \langle \sigma^+_{n}\sigma^-_{m} \rangle
	\biggr)
	\end{split}
	\label{eq:c3f}\\
	\begin{split}
		\partial_t \langle \sigma^z_{n}\sigma^+_{m}\sigma^-_{k} \rangle &=
		- 2\Gamma_T \langle \sigma^z_{n}\sigma^+_{m}\sigma^-_{k} \rangle
  +\Gamma_\Delta \langle \sigma^+_{n}\sigma^-_{m} \rangle 
		- 8 g \Im\!\left[\langle \sigma^+_{0}\sigma^-_{n} \rangle\right]\langle \sigma^+_{n}\sigma^-_{m} \rangle \\
		&+2g  \biggl(
\Im\!\left[\langle \sigma^+_{0}\sigma^-_{n} \rangle\right] \langle \sigma^z_{n}\sigma^z_{m} \rangle -
2\Im\!\left[ \langle \sigma^+_{0}\sigma^-_{n} \rangle\right]\langle \sigma^z_{n} \rangle^2 +
2\Im\!\left[\langle \sigma^+_{0}\sigma^-_{n}\sigma^z_{m} \rangle\right] \langle \sigma^z_{n} \rangle
		\biggr)
	\end{split}\label{eq:c3g}\\
	\partial_t \langle \sigma^z_{n}\sigma^z_{m} \rangle &=
 - 2\Gamma_T\langle \sigma^z_{n}\sigma^z_{m} \rangle +
2\Gamma_\Delta\langle \sigma^z_{n} \rangle
- 8 g\Im\!\left[\langle \sigma^+_{0}\sigma^-_{n}\sigma^z_{m} \rangle\right]
	\label{eq:c3h} \\
	\begin{split}
		\partial_t \langle \sigma^z_{0}\sigma^z_{n} \rangle &=
		- (\kappa+\Gamma_T) \langle \sigma^z_{0}\sigma^z_{n} \rangle - \kappa \langle \sigma^z_{n} \rangle + \Gamma_\Delta \langle \sigma^z_{0} \rangle
		+4g(N-1) \Im\!\left[\langle \sigma^+_{0}\sigma^-_{n}\sigma^z_{m} \rangle\right]
	\end{split}\label{eq:c3i}
\end{align}
 In writing
\cref{eq:c3e,eq:c3f,eq:c3g} fourth-order moments were approximated by setting the 
fourth-order cumulants to zero:
\begin{align}
\langle \langle \sigma^+_{0}\sigma^-_{n} \sigma^+_{m}\sigma^-_{k} \rangle \rangle=0,\quad 
\langle \langle \sigma^z_{0}\sigma^z_{n} \sigma^+_{m}\sigma^-_{k} \rangle \rangle=0,\quad 
\langle \langle \sigma^+_{0}\sigma^-_{n} \sigma^z_{m}\sigma^z_{k} \rangle \rangle=0,
\end{align}
where
 \begin{align}
\begin{gathered}
\langle\langle \sigma_{a}^{\alpha}\sigma_{b}^{\beta}\sigma_{c}^{\gamma}\sigma_{d}^{\delta}\rangle\rangle := 
\langle \sigma_{a}^{\alpha}\sigma_{b}^{\beta}\sigma_{c}^{\gamma}\sigma_{d}^{\delta}\rangle 
 - \langle \sigma_{a}^{\alpha}\sigma_{b}^{\beta}\rangle\langle \sigma_{c}^{\gamma}\sigma_{d}^{\delta}\rangle - \langle \sigma_{a}^{\alpha}\sigma_{c}^{\gamma}\rangle\langle \sigma_{b}^{\beta}\sigma_{d}^{\delta}\rangle 
  - \langle \sigma_{a}^{\alpha}\sigma_{d}^{\delta}\rangle\langle \sigma_{b}^{\beta}\sigma_{c}^{\gamma}\rangle
 \\
-\langle \sigma_{a}^{\alpha}\rangle\langle \sigma_{b}^{\beta}\sigma_{c}^{\gamma}\sigma_{d}^{\delta}\rangle 
- \langle \sigma_{b}^{\beta}\rangle \langle \sigma_{a}^{\alpha}\sigma_{c}^{\gamma}\sigma_{d}^{\delta}\rangle
 - \langle \sigma_{a}^{\alpha}\sigma_{b}^{\beta}\sigma_{d}^{\delta}\rangle\langle \sigma_{c}^{\gamma}\rangle
 - \langle \sigma_{a}^{\alpha}\sigma_{b}^{\beta}\sigma_{c}^{\gamma}\rangle\langle \sigma_{d}^{\delta}\rangle
\\
+2\langle \sigma_{a}^{\alpha}\rangle\langle \sigma_{b}^{\beta}\rangle\langle \sigma_{c}^{\gamma}\sigma_{d}^{\delta}\rangle
 + 2\langle \sigma_{a}^{\alpha}\rangle\langle \sigma_{b}^{\beta}\sigma_{c}^{\gamma}\rangle\langle \sigma_{d}^{\delta}\rangle + 2\langle \sigma_{a}^{\alpha}\rangle\langle \sigma_{b}^{\beta}\sigma_{d}^{\delta}\rangle\langle \sigma_{c}^{\gamma}\rangle \\
 + 2\langle \sigma_{a}^{\alpha}\sigma_{b}^{\beta}\rangle\langle \sigma_{c}^{\gamma}\rangle\langle \sigma_{d}^{\delta}\rangle
 + 2\langle \sigma_{a}^{\alpha}\sigma_{c}^{\gamma}\rangle\langle \sigma_{b}^{\beta}\rangle\langle \sigma_{d}^{\delta}\rangle + 2\langle \sigma_{a}^{\alpha}\sigma_{d}^{\delta}\rangle\langle \sigma_{b}^{\beta}\rangle\langle \sigma_{c}^{\gamma}\rangle 
 \\
 - 6\langle \sigma_{a}^{\alpha}\rangle\langle \sigma_{b}^{\beta}\rangle\langle \sigma_{c}^{\gamma}\rangle\langle \sigma_{d}^{\delta}\rangle.\label{eq:4cumulant}
\end{gathered}
 \end{align}
 Note that many of these terms vanish for the model with \(\text{U}(1)\) symmetry.
\end{widetext}

 \appsection{Behavior of correlations as \texorpdfstring{\(N\to\infty\)}{N tends to infinity}}
 \label{app:2}

 \begin{figure}[h]
	\centering
    \vspace{-2\baselineskip}%
	\phantomsubfloat{\label{fig:sm2a}}%
    \phantomsubfloat{\label{fig:sm2b}}%
    \phantomsubfloat{\label{fig:sm2c}}%
    \phantomsubfloat{\label{fig:sm2d}}%

	\hspace{-.25cm}%
	\includegraphics[width=1.025\linewidth]{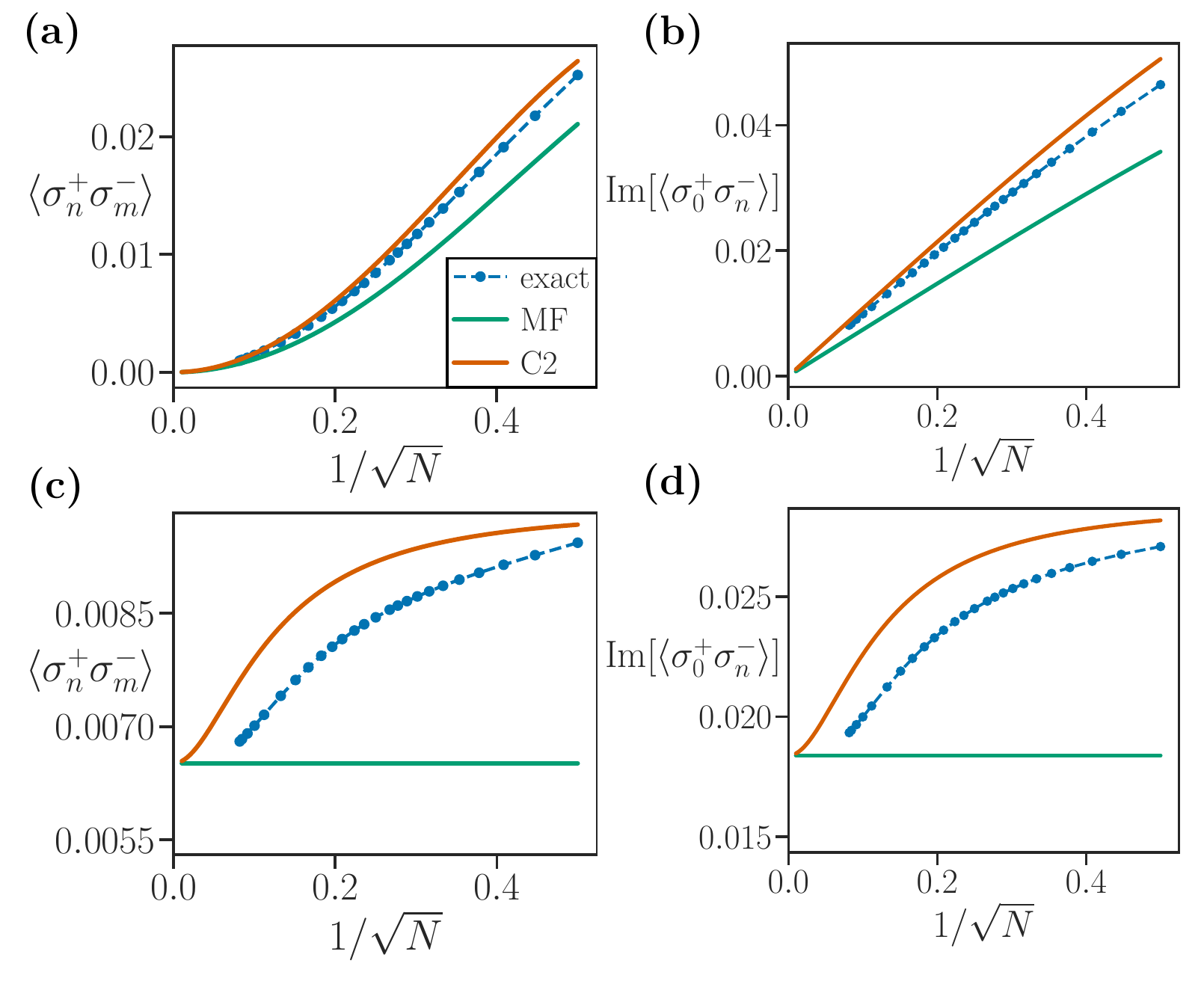}%
 \vspace{-\baselineskip}
	\caption{%
		(a) Satellite-satellite and (b) central-satellite correlations
  in the steady state of the model with scaling \(g\sqrt{N}\) fixed
  and parameters as in \cref{fig:2a} (\(g\sqrt{N}=3\), \(\kappa=1\)).
  Exact data (blue dots) up to \(N=150\) and second-order (C2) results for the correlations 
  are included, as well as the mean-field (MF) approximations
  \(\langle \sigma^+_n \sigma^-_m\rangle\approx\left\lvert\langle \sigma^+_n \rangle\right\rvert^2 \) and
   \(\langle \sigma^+_0 \sigma^-_n\rangle
   \approx
   \langle \sigma^+_0 \rangle
   \langle \sigma^-_n \rangle
   \).
  Note the use of \(1/\sqrt{N}\) on the horizontal axis: for this
  scaling \(\langle \sigma^+_n \sigma^-_m\rangle=o(1/\sqrt{N})\)
  and \(\Im\!\left[\langle \sigma^+_0 \sigma^-_n\rangle\right]=O(1/\sqrt{N})\)
  as \(N\to\infty\) (the real part of \(\langle \sigma^+_0 \sigma^-_n\rangle\) vanishes at resonance).
    (c), (d) Same correlations when instead \(\kappa/N\)
    is fixed with parameters as in \cref{fig:2b}
    (\(\kappa/N=1/16\), \(g=3/4\)). In this case both pairs
    of correlations remain finite for all \(N\).
	}
	\label{fig:sm2}
\end{figure}
 
In this appendix we show the behavior of pairwise correlations as \(N\to\infty\)
for the central spin model.  These results support the arguments for convergence
made in \cref{sub:fixed_kappa}.

\Cref{fig:sm2a,fig:sm2b} show satellite-satellite  \(\langle \sigma^+_n \sigma^-_m\rangle\)
and central-satellite
 \(\langle \sigma^+_0 \sigma^-_n\rangle\)
correlations against \(1/\sqrt{N}\) for the model at fixed
\(g\sqrt{N}\). We show exact results
up to \(N=150\) as well as the prediction of second-order cumulants
and mean-field theory. Notice in particular that 
 \(\langle \sigma^+_n \sigma^-_m\rangle\) decays faster than
 \(1/\sqrt{N}\) as \(N\to\infty\)
 (vanishing gradient at $1/\sqrt{N} \to 0$ in \cref{fig:sm2a}). As such, at large \(N\), the terms
 \(\sim g \langle \sigma^z_0\rangle,\sim g \langle \sigma^z_n\rangle \) present in the second-order
equation \cref{eq:c2a} (but not mean-field theory) are dominant compared to 
the final term  \(\sim g \langle \sigma^z_0\rangle\langle\sigma^+_n \sigma^-_m\rangle  \) occurring there. 

When instead considering the correlations at fixed \(\kappa/N\), shown in 
\cref{fig:sm2c,fig:sm2d}, we see that both tend to finite limits,
allowing for the reduction of the second-order cumulant equations to
mean-field theory when \(N\to\infty\) as argued in the main text.

Thus we have both a case where correlations vanish as
\(N\to\infty\) but mean-field and second-order cumulants do
not have a well-defined limit [\cref{fig:2a}],
and a case where they remain finite yet the two approaches
have a common limit capturing the exact behavior [\cref{fig:2b}].
This makes evident the fact that knowledge of the behavior of
correlations as \(N\to\infty\) is not sufficient to conclude
the correctness of mean-field theory or the convergence of
higher-order cumulant expansions in this limit.

 \appsection{Third-order cumulant equations with symmetry breaking}
 \label{app:3}

 \begin{figure}[h]
	\centering
    \vspace{-2\baselineskip}%
	\phantomsubfloat{\label{fig:sm1a}}%
    \phantomsubfloat{\label{fig:sm1b}}%
    \phantomsubfloat{\label{fig:sm1c}}%
    \phantomsubfloat{\label{fig:sm1d}}%

	\hspace{-.2cm}%
	\includegraphics[width=\linewidth]{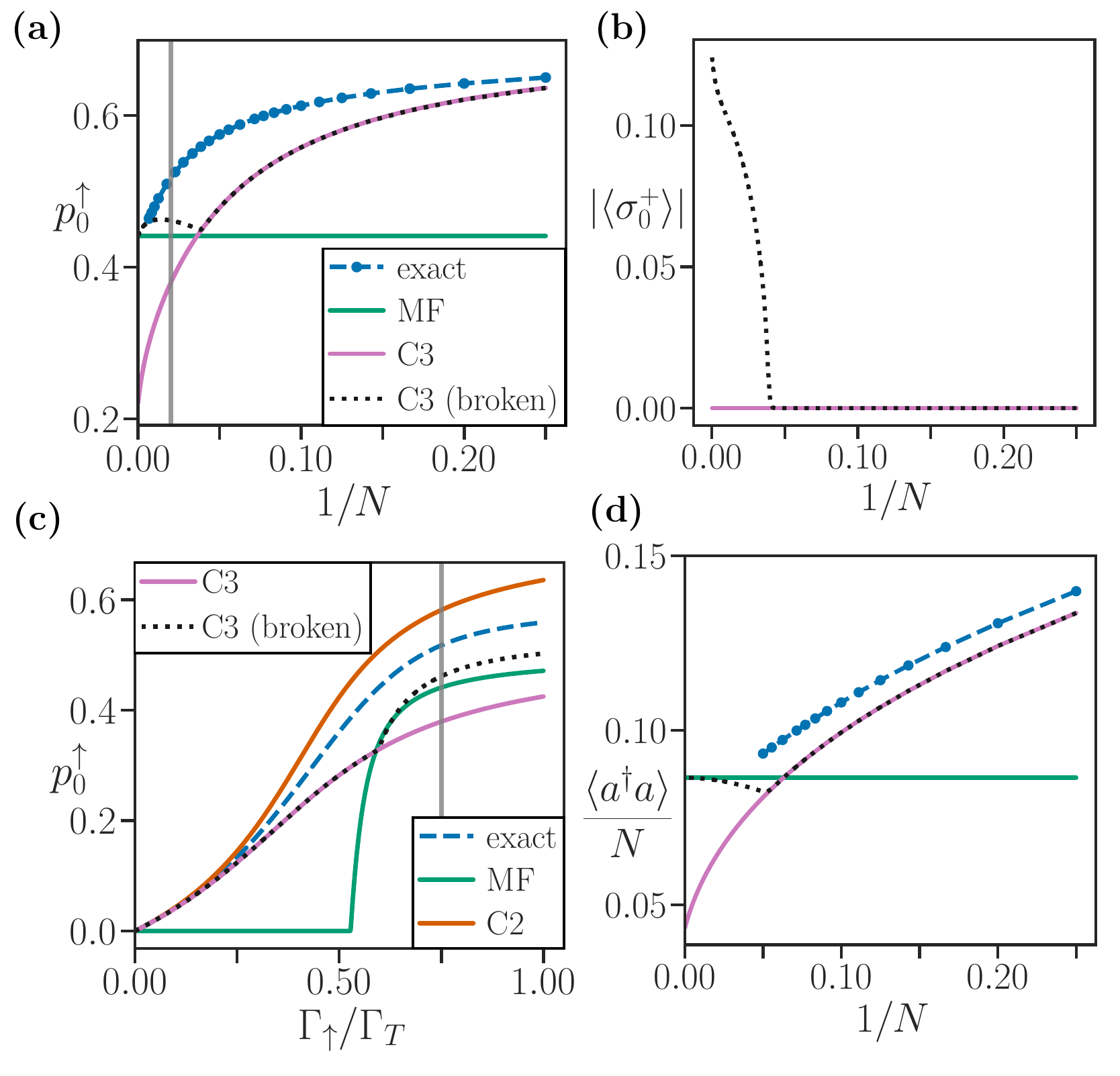}%
 \vspace{-\baselineskip}
	\caption{%
		(a) Exact, mean-field (MF), and third-order (C3) results
  for the steady state central-site population \(p_0^\uparrow=[1+\langle \sigma^z_0\rangle]/2\)
  of the central spin model at fixed \(\kappa/N\)
  [parameters as in \cref{fig:2b}:  \(\kappa/N=1/16\), \(g=3/4\)].
  Third-order results retaining symmetry-breaking terms in the equations
  are indicated with a dotted black line. (b) At \(N=26\) 
  symmetry breaking \(\langle \sigma^+_0\rangle \neq 0\) occurs
  in the steady state of these equations corresponding to the turning point of
  the dotted line in (a). (c) Central population versus
  \(\Gamma_\uparrow/\Gamma_T\) (\(\Gamma_T=\Gamma_\uparrow+\Gamma_\downarrow\)) at \(N=50\) 
 as in   \cref{fig:2c} (\(\kappa/N=1/16\), \(g=3/4\)), now including
  third-order results with and without symmetry-breaking terms. The gray
  vertical line indicates data from (a).  
  (d) Exact, mean-field (MF), and third-order (C3) results
  for the scaled photon number in the Tavis--Cummings model
  with parameters as in \cref{fig:3b} (\(g\sqrt{N}=9/10\), \(\Gamma_\uparrow=3\Gamma_T/4\)).
	}
	\label{fig:sm1}
\end{figure}
 
 In this appendix we provide the results of third-order cumulant expansions
 with symmetry-breaking terms for the central spin and Tavis--Cummings models.
 
 Retaining moments, e.g., 
 \(\langle \sigma^+_0 \sigma^+_n \rangle \),
 in the equations of motion that would otherwise 
 vanish under \(\text{U}(1)\)  symmetry significantly
 increases the number of equations required 
 to form  a complete set at any given order. We used
 the QuantumCumulants.jl Julia package~\cite{plankensteiner2022}
 to derive the third-order
 equations in each case. Using an initial state 
that breaks the symmetry, these equations
 were evolved to long times to obtain a numerical
 approximation of the steady state.

Since the coefficients of terms in the definition of a cumulant 
always sum to zero [cf. \cref{eq:4cumulant}], when one sets
a cumulant to zero to obtain an approximation for a high-order
moment, the number of terms in the approximation for that moment,
accounting for their signs, is 1. That is, provided no terms in
the cumulant vanish due to symmetry considerations.
As a result, in the presence of symmetry breaking there is
no longer disparity between the
asymptotic form of odd-order cumulant equations and mean-field theory
as \(N\to\infty\) due to the factorization of
moments \(\langle \sigma^+_0 \sigma^-_n \sigma^+_m \sigma^-_k\ldots \rangle\). 

In line with the above, \cref{fig:sm1a} shows a common \(N\to\infty\) limit for the third-order equations with symmetry breaking (dotted line) and mean-field theory. Note however there is a range of \(N\)
[\(N \leq 26\) in \cref{fig:sm1a}] for which symmetry breaking is not
present in the obtained steady state (\cref{fig:sm1b}) and
the original third-order results are followed by the dotted line.
Further, even with symmetry breaking the third-order results cannot
be relied upon to provide a better
approximation than a second-order expansion. This is clearly seen in
\cref{fig:sm1c}, which shows \(p^\uparrow_0\) against \(\Gamma_\uparrow/\Gamma_\downarrow\) at \(N=50\). We point out the
agreement of all cumulant expansions at pump strengths well below the mean-field
threshold, where $p_0^\uparrow$ must vanish as \(N\to\infty\).  Note also the crossing of the third-order (symmetry-preserving) and mean-field curves which marks the transition to
the symmetry-broken steady state;  this is inevitable at large $N$, where the third-order result is below the mean-field prediction.

Finally, in \cref{fig:sm1d} we observe similar behavior with the 
third-order equations with symmetry-breaking terms for the Tavis--Cummings model, although in this case the mean-field limit is approached
from below.

\bibliography{refs.bib}
\end{document}